\documentclass[aps,prl,twocolumn,showpacs,superscriptaddress,letterpaper]{revtex4}
\usepackage{graphicx,epsfig,amsmath,amssymb,amsfonts,mathrsfs,psfrag,color}
\usepackage{bm,txfonts,mathrsfs,mathtools,relsize}
\newcommand{\mat}[4]{\left(\begin{array}{cc}#1&#2\\#3&#4\end{array}\right)}

\begin{document} 

\title{Neutral mode heat transport and fractional quantum Hall shot noise }
\author{So Takei}
\affiliation{Max-Planck-Institut f\"ur Festk\"orperforschung, D-70569 Stuttgart, Germany}
\affiliation{Condensed Matter Theory Center, Department of Physics, The University of Maryland College Park, MD 20742}
\author{Bernd Rosenow}
\affiliation{Max-Planck-Institut f\"ur Festk\"orperforschung, D-70569 Stuttgart, Germany}
\affiliation{Institut f\"ur Theoretische Physik, Universit\"at Leipzig, D-04103, Leipzig, Germany}
\date{\today}
\pacs{73.43.Jn,73.43.Cd,73.50.Td,71.10.Pm}

\begin{abstract}

We study nonequilibrium edge state transport in the fractional quantum Hall regime for states with one or several counter-propagating 
neutral modes. We consider a setup in which the neutral modes are heated by a hot spot, and where heat transported by the neutral 
modes causes a temperature difference between the upper and lower edges in a Hall bar. This temperature difference is probed by the 
excess noise it causes for scattering across a quantum point contact. We find that the excess noise in the quantum point contact provides 
evidence for counter-propagating neutral modes, and we calculate its dependence on both the temperature difference between the edges 
and on source drain bias.
\end{abstract}

\maketitle

Many of the peculiar properties of quantum Hall (QH) systems can be attributed to the existence of quasi-one-dimensional 
electronic states along the perimeter  of the sample, the so-called edge states \cite{halperin}. In the integer quantum 
Hall regime edge states can be modelled by non-interacting electrons, and the physics of edge states is capable of 
describing numerous transport experiments if the Landauer transport theory is generalized to incorporate multiple 
terminals \cite{buettiker}. In the fractional QH regime interactions play an essential role, and edge states 
must be described as Luttinger liquids \cite{wen}, in some cases with excitations propagating both with and against 
the orientation imposed by the magnetic field. For instance, in the case of filling fraction $\nu=2/3$ two 
counter-propagating edge modes are predicted \cite{wen,macdonald}, which would give rise to non-universal Hall and 
two-terminal conductances. Experimentally, however, conductances are quantized and a counter-propagating charge 
mode was not observed \cite{ashoorietal}. This problem is resolved by taking into account that in the presence of random 
edge scattering the $\nu=2/3$ edge undergoes reconstruction into a disorder-dominated phase with a single 
downstream-propagating charge mode and a single upstream-propagating neutral mode \cite{kfp}. 

Interest in neutral quantum Hall edge modes was revived because one or several neutral Majorana edge mode 
is expected to encode the non-abelian statistics of the QH state at filling fraction $\nu=5/2$ \cite{review,MR,apf1,apf2,OvWe08}. 
Neutral quantum Hall modes are notoriously difficult to observe as they do not participate in charge transport. Recently, 
experimental evidence for neutral modes was presented by demonstrating that injection of a DC current can influence the 
low frequency noise generated at a quantum point contact (QPC) located {\em upstream} of the contact where the current 
is injected \cite{heiblum}. Partitioning of a DC current by a QPC and the influence of downstream heat transport on a second 
QPC was studied both experimentally \cite{Granger+09} and theoretically \cite{FeLi08,GrDa09}. 

In this Letter, we theoretically analyze a setup akin to that of Ref. \cite{heiblum} and find that a current injected into a quantum 
Hall mode downstream of a QPC indeed enhances the charge noise due to scattering at the QPC. In our model, this happens 
because the injected current causes a hot spot in the contact and, in the presence of one or several neutral modes propagating in 
the direction opposite to that imposed by the magnetic field for charge pulses, heat is conducted from the contact to the QPC and 
gives rise to excess noise in the current scattered across the QPC. When the model is generalized to the non-abelian $\nu=5/2$ 
quantum Hall state the enhancement of the charge noise, also observed in this state \cite{heiblum}, limits the possible descriptions of 
the state to those that support counter-propagating neutral modes, namely, the anti-Pfaffian \cite{apf1,apf2} and an edge reconstrcted 
Pfaffian state \cite{Wan+06,OvWe08}.
\begin{figure}[t]
\begin{center}
\includegraphics[scale=0.6]{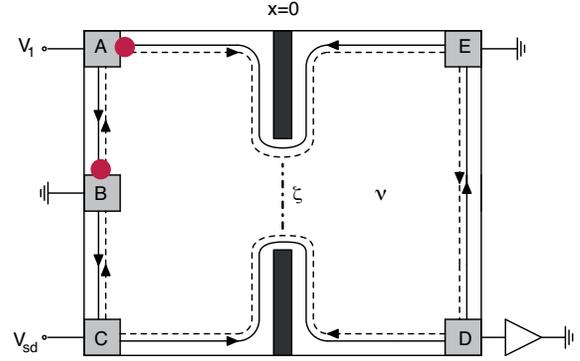}
\caption{\label{fig:sys} (color online) Schematic diagram of the considered setup.  The magnetic field is oriented so that charge transport 
along the edge (solid black lines) occurs with the counterclockwise chirality. There is one or several neutral modes 
(dashed lines) propagating clockwise, and the dash-dotted line denotes backscattering at the constriction. Hot spots caused by a finite 
$V_1$  are indicated by red dots.}
\end{center}
\end{figure}

We consider a multi-terminal Hall bar geometry (see Fig.\ref{fig:sys}), and we assume that 
the bulk is in a quantized Hall state with filling fraction $\nu$ with one or several neutral modes propagating in the direction 
opposite to that of the charge mode. The two edges between which scattering takes place are located on the upper and lower sides 
of the Hall bar, and are labeled $j=1$ and $j=2$, respectively. While contacts $B$ and $E$ are grounded, contacts $A$ and $C$ have 
tuneable electrochemical potentials, $eV_1$ and $eV_{\rm sd}$, where $e$ here is the electron charge. Current and noise are measured 
at contact $D$. The two dark pads at $x=0$ represent top gates, which form a gated constriction that pinches the edge channels 
together and causes backscattering of quasiparticles (QPs) between the edges. 

The source-drain bias at contact $C$ raises the chemical potential of the emanating charge mode in the bottom edge and gives 
rise to a current $\nu (e^2/h) V_{sd}$ impinging on the QPC. A finite $V_{1}$   gives rise to an electrical current 
$I_1 = \nu (e^2/h) V_1$ flowing from contact $A$ to contact $B$. We note that in the experimentally relevant regime where Hall 
and longitudinal conductances are quantized, no electrical current is flowing from contact $A$ to contact $E$, and that the expectation 
value of the neutral mode decays quickly away from contact $A$. While the total electrical power supplied to the system is $I_1 V_1$, 
the electrical energy current flowing from contact $A$ to contact $B$, which is dissipated in a hot spot at  contact $B$, is only $I_1 V_1/2$ 
\cite{ChHa}. The rest of the electical power is dissipated at a second hot spot, located on the upstream side of contact $A$. Here, high 
energy electrons ``fall into" the incoming edge mode and fill it up to the electrochemical potential $e V_1$ of contact $A$, dissipating energy 
in the process. The heat generated in this process has to be transported away, which may happen through the wire connecting to contact $A$ 
or some other cooling mechanism. In general, the equilibrium temperature $T_A$ in the region of the hot spot will grow monotonically with 
the current $I_1$. Under the specific assumption that the cooling mechanism is of electronic origin and follows the Wiedemann-Franz law, 
one would find that the temperature $T_A$ at contact $A$ is given by $T_A = \sqrt{T_0^2 + I_1 V_1 /G L}$, where $T_0$ denotes the 
temperature of the electron bath contact $A$ is connected to, $G$ the conductance of contact $A$ to that electron bath, and 
$L$ the Lorenz number. In the limit where $T_A \gg T_0$, one finds $T_A \propto I_1$. 

If the upper edge has at least one counter-propagating neutral mode, heat transport from contact $A$ to the QPC will be possible and the 
hot spot at contact $A$ will give rise to an increased temperature $T_1 $ of the upper edge at the QPC. Denoting the temperature of the 
lower edge by $T_2$, the fact that $T_1 > T_2$ due to injection of a current $I_1$ into contact $A$ gives rise to enhanced scattering at 
the QPC and an enhancement of current noise. This description is justified because on the scale of the inelastic mean free path $\ell_\sigma$ 
equilibration between the charge mode and neutral mode(s) takes place \cite{kfnm}. If the distance between  contact $A$ and the QPC  is 
much larger than $\ell_\sigma$, charge and neutral modes have a common temperature $T_1$ at the QPC. In addition, we make the realistic 
assumption (verified for the random 2/3-edge) that $\ell_\sigma \gg L_T$, where $L_T = u_\sigma/T$ denotes the thermal length. As the 
edge correlations describing scattering at the QPC decay on the scale $L_T$, the inequality $L_T \ll \ell_\sigma$ implies that a possible 
temperature gradient on the scale $\ell_\sigma$ will not influence current and noise at the QPC, and we can consider an effective model 
in which backscattering at the constriction is described by assigning a common temperature $T_1$ to both charge and neutral modes on 
the upper edge. The relation between the temperatures $T_1$ and $T_A$ depends on the thermal Hall conductance $K_H$ of the edge. 
For a vanishing $K_H = 0$ (realized for a random 2/3-edge), heat transport along the edge is diffusive, and $T_1 < T_A$. The exact value 
of $T_1$ depends on microscopic details like the distances between the QPC to contacts $A$ and $E$ and the amount of scattering between 
different edge modes. For a $K_H < 0$ (e.g. for the anti-Pfaffian edge \cite{apf1,apf2} one finds $K_H = - 1/2$), heat transport is ballistic 
and $T_1 = T_A$. In the following, we present a calculation for current and noise at contact $D$ as a function of both source-drain 
voltage $V_{sd}$ and ``neutral" voltage $V_1$ for the random $2/3$-edge. We later generalize our formulas to account for general states.

In the presence of disorder, edge excitations of the $\nu=2/3$ fractional QH liquid are predicted to reflect 
the physics of a stable zero-temperature disorder-dominated fixed point \cite{kfp,kfnm}. At the fixed point, each edge consists of a set of 
decoupled charge ($\phi_\rho$) and neutral ($\phi_\sigma$) modes that propagate in opposite directions. The effect of random elastic scattering 
can be incorporated into the neutral mode by fermionizing it, eliminating the scattering term by a spatially random SU(2) transformation, and 
rebosonizing. At the fixed point, the appropriate real-time Lagrangian density is given by 
$\mathscr{L}_0=\sum_{j=1,2}(\mathscr{L}_{\rho j}+\mathscr{L}_{\sigma j})$,  where
%
\begin{eqnarray}
\mathscr{L}_{\rho j}&=&\partial_x\phi_{\rho j}((-)^{j-1}\partial_t-u_\rho\partial_x)
\phi_{\rho j}/2\nu\nonumber \\ 
\mathscr{L}_{\sigma j}&=&\partial_x\phi_{\sigma j}((-)^j\partial_t-u_\sigma\partial_x)
\phi_{\sigma j}/4 \ \ . \label{fixedpoint.eq}
\end{eqnarray}
%
Here, $u_\rho$ ($u_\sigma$) is the charge (neutral) mode velocity, and we use units where $\hbar=1=k_B$. The charge and neutral modes 
are coupled by a spatially random  interaction term $\mathscr{L}_{\rho \sigma} = u_{int}(x) \partial_x \phi_{\rho j} \partial_x \phi_{\sigma j}$.
The coupling $u_{int}(x)$ is uncorrelated on spatial scales large compared to the elastic mean free path $\ell_0$, and we denote its variance by 
$W_{int}$. This term decays under the renormalization group (RG) flow \cite{kfnm} and vanishes in the zero temperature limit, giving rise to 
the fixed point Lagrangian Eq.~(\ref{fixedpoint.eq}). At finite temperature, the RG flow is stopped at the thermal length $L_T$, and the 
coupling between the charge and neutral modes gives rise to an inelastic mean free path $\ell_\sigma^{-1}\sim W_{\rm int}T^2$ \cite{kfnm}. 
At low temperatures $\ell_\sigma$ is parametrically larger than the thermal length $L_T$ over which the bosonic Green function decays. Hence, 
the local bosonic expectation values needed to evaluate the probability of QP scattering across the QPC can be evaluated using the fixed point 
Lagrangian Eq.~(\ref{fixedpoint.eq}). 

\begin{figure}[t]
\begin{center}
\includegraphics[scale=0.52]{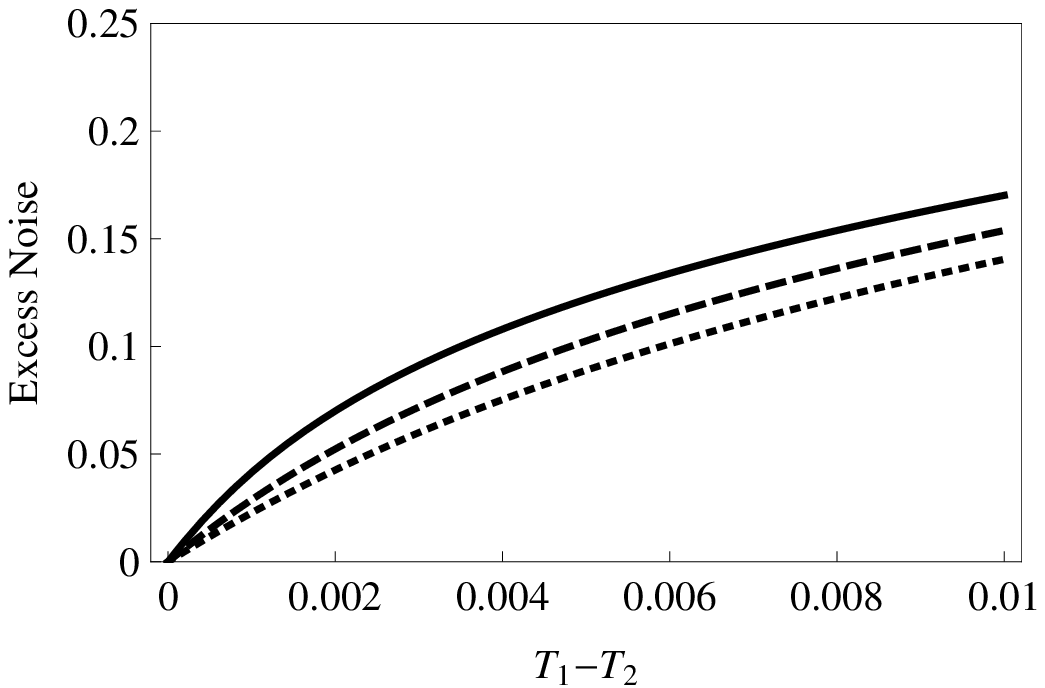}
\includegraphics[scale=0.5]{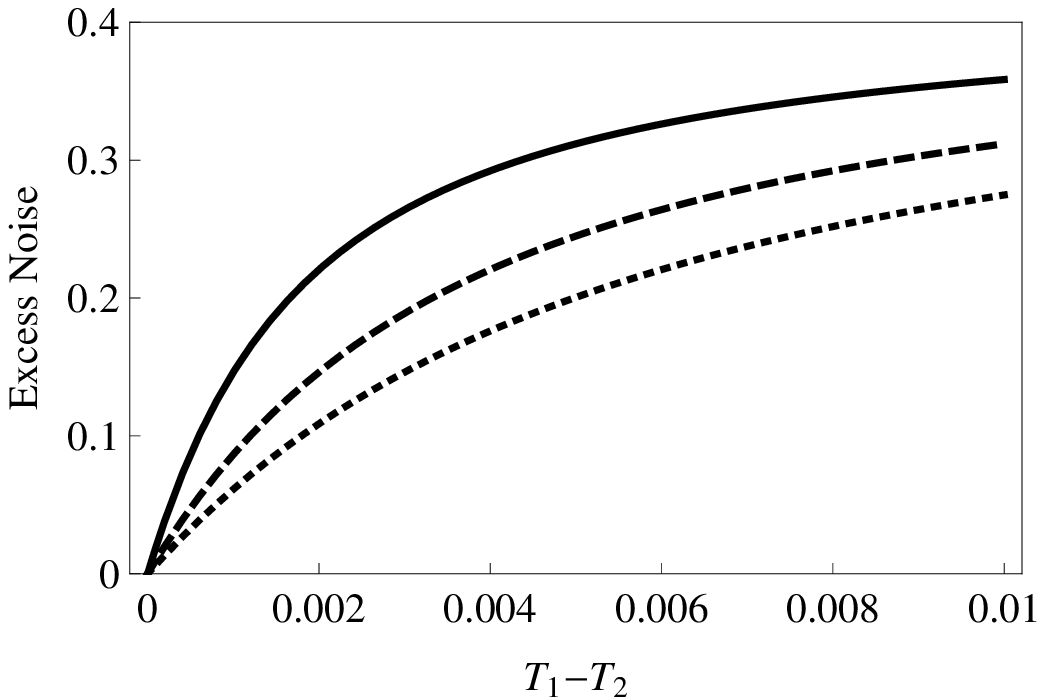}
\caption{\label{fig:noise} Nonequilibrium excess noise for the random  $\nu=2/3$ edge (upper panel) and the random $\nu=5/2$ anti-Pfaffian 
edge (lower panel) as a function of the temperature difference, $\Delta T=T_1-T_2$, between the two edges. The solid, dashed and dotted 
curves are for $T_2=0.001$, $T_2=0.002$, and $T_2=0.003$, respectively. Tunneling amplitudes and temperature are in units of the UV cutoff $\tau_c^{-1}$, and the noise is in units of the Nyquist noise with $T_2=0.001$. Tunneling amplitudes have been chosen so that the QPC has a 
transmission of 0.9 at $T_1=T_2=0.001$.}
\end{center}
\end{figure}
We now outline the formalism which enables us to compute the current and noise. Upon integrating out all fluctuations away from the defect 
site (at $x=0$) in the action $S_0 = \int dt \mathscr{L}_0$, we arrive at an effective action in terms of the local fields 
$q_{\rho j}(t):=\phi_{\rho j}(x=0,t)$ and $q_{\sigma j}(t):=\phi_{\sigma j}(x=0,t)$
\cite{kfnoise} 
\begin{eqnarray}
\label{SeffQ}
S^K_{{\rm eff}}&=&\frac{i}{4\pi}\sum_{m=\rho,\sigma}\sum_{j=1,2}\int\frac{d\omega}{2\pi}
Q_{mj}^\dagger(\omega)\hat d_{mj}^{-1}(\omega)Q_{mj}(\omega)\nonumber\\
&+&\int\frac{d\omega}{\pi}\Gamma^\dagger(\omega)\mat{0}{\frac{\omega}{2\pi}}{\frac{\omega}{2\pi}}
{\frac{\omega}{2\pi}\coth\left(\frac{\omega}{2T_2}\right)}Q_{\rho2}(\omega)
\nonumber\\
&+&\int\frac{d\omega}{2\pi}\,\Gamma^\dagger(\omega)\mat{0}{-\frac{\omega\nu}{4\pi}}
{\frac{\omega\nu}{4\pi}}{-\frac{\omega\nu}{4\pi}\coth\left(\frac{\omega}{2T_2}\right)}\Gamma(\omega).
\end{eqnarray}
To harness the nonequilibrium nature of the problem, the above action has been mapped onto the Keldysh time-loop contour 
\cite{kamenev,rammerbook}. Upper case letters are used to denote two-component fields in Keldysh space, i.e. $B=(b^{\rm cl},\, b^{\rm q})^T$ 
for a general bosonic field $b$. The components are labeled ``classical" and ``quantum", which relate to the fields on the forward ($+$) and 
backward ($-$) branches of the Keldysh contour via $b^{\rm cl, q}=(b^+\pm b^-)/\sqrt{2}$. $\hat d_{mj}(\omega)$ is the local Keldysh matrix 
propagator for mode $m\in\{\rho,\sigma\}$ and edge $j\in\{1,2\}$. Each propagator has the Keldysh causality structure \cite{kamenev}, and 
contains retarded ($R$), advanced ($A$) and Keldysh ($K$) Green's functions. Here, the retarded Green's functions are given by 
$d^R_{\rho j}(\omega)=[d^A_{\rho j}(\omega)]^*=-i\nu/2\omega$ and $d^R_{\sigma j}(\omega)=[d^A_{\sigma j}(\omega)]^*=-i/\omega$. 
The Keldysh Green's functions can be obtained via the fluctuation-dissipation relation, 
$d^K_{mj}(\omega)=\coth(\omega/2T_j)(d^R_{mj}(\omega)-d^A_{mj}(\omega))$. In Eq.(\ref{SeffQ}), we have also introduced 
$\Gamma(\omega)=(b(\omega),\, e\mu(\omega))^T$. Its classical component, $b(\omega)$, is related to the external source-drain voltage 
through $\partial_tb(t)=\sqrt{2}eV_{\rm sd}$. Its quantum component is the source field, $\mu(\omega)$, which is used to generate all the 
cumulants of the current operator defined on the lower edge at position $x_0>0$, i.e. $I(x_0,t)=eu_\rho\partial_x\phi_{\rho 2}(x_0,t)/\sqrt{2\pi}$.
In the above, we have assumed that the period of the AC source-drain bias is much longer than the time for ballistic transport through the device, 
thus, effectively allowing one to take the limit $x_0\rightarrow 0^+$. The limit entails no effect on our results which only focus on the steady steady 
current and the low frequency noise.

\begin{figure}[t]
\begin{center}
\includegraphics[scale=0.52]{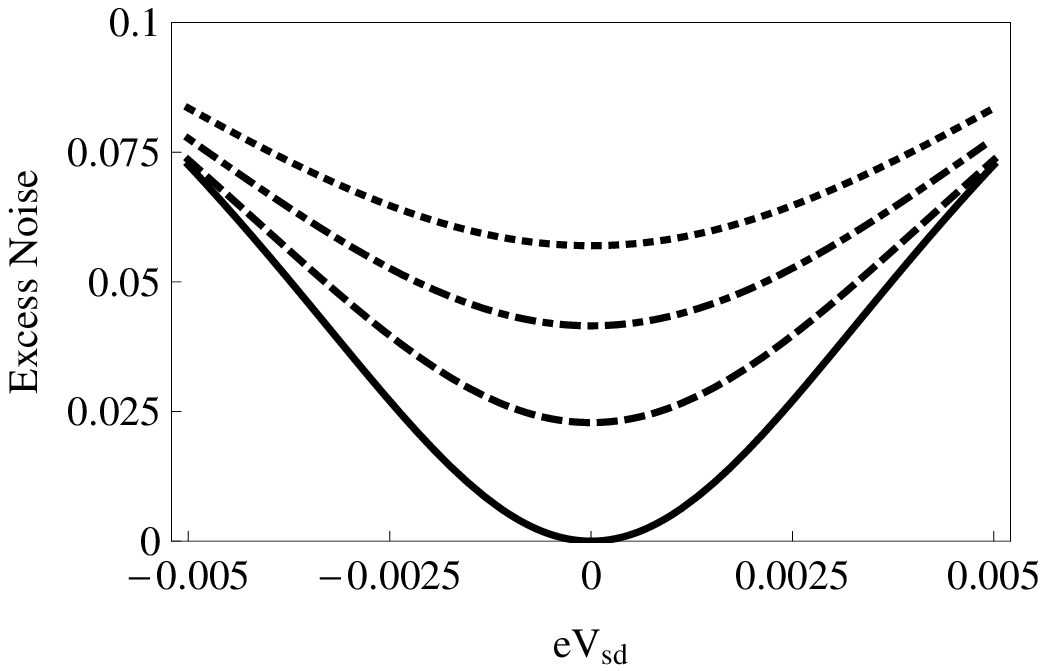}
\includegraphics[scale=0.5]{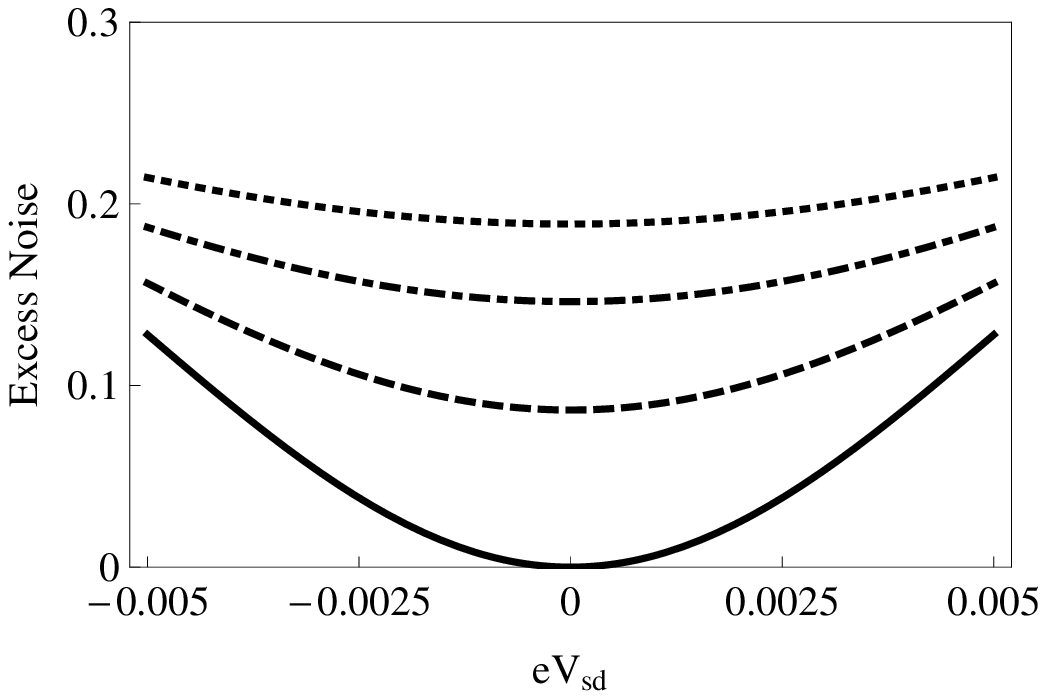}
\caption{\label{fig:noisebias} Nonequilibrium excess noise for the random $\nu=2/3$ edge (upper panel) and the random $\nu=5/2$ anti-Pfaffian 
edge (lower panel) as a function of the source-drain voltage. The bottom edge temperature is fixed at $T_2=0.001$, and the top edge temperature 
$T_1=0.001$ (Solid); $T_1=0.0015$ (Dashed); $T_1=0.002$ (Dot-dashed); and $T_1=0.0025$ (Dotted).  Tunneling amplitudes and temperature 
are in units of the UV cutoff $\tau_c^{-1}$, and the noise is in units of the Nyquist noise with $T_2=0.001$. Tunneling amplitudes have been chosen 
so that the QPC has a transmission of 0.9 at $T_1=T_2=0.001$.}
\end{center}
\end{figure}
For $\nu=2/3$, the most relevant operator which describes QP tunneling between the edges is not unique. In particular, there are three tunneling 
terms with the same scaling dimension, two of which involve tunneling of $e/3$ QPs and another involving $2e/3$ QPs. Since tunneling takes place 
at $x=0$, the tunneling Lagrangians can be expressed in terms of the local fields. On the Keldysh contour, the action is given by 
$S^K_T=-i\int dt\,[L^+_T-L^-_T]$, where
\begin{eqnarray}
\label{tunnelS}
L^\alpha_T&=&-\zeta_1\cos\left[\left(q^\alpha_{\rho 1}(t)-q^\alpha_{\sigma 1}(t)
-q^\alpha_{\rho 2}(t)+q^\alpha_{\sigma 2}(t)\right)/2\right]\nonumber\\
&&-\zeta_2\cos\left[\left(q^\alpha_{\rho 1}(t)+q^\alpha_{\sigma 1}(t)
-q^\alpha_{\rho 2}(t)-q^\alpha_{\sigma 2}(t)\right)/2\right]\\
&&-\zeta_3\cos\left(q^\alpha_{\rho 1}(t)-q^\alpha_{\rho 2}(t)\right),\nonumber
\end{eqnarray}
and $\zeta_i$ are the tunneling amplitudes. Here, $\alpha\in\{+,-\}$ labels the forward and backward branches of the Keldysh contour.

The terms linear in $\Gamma(\omega)$ in Eq.(\ref{SeffQ}) can be eliminated by performing the shift
\begin{equation}
\label{qshift}
Q_{\rho2}(\omega)\rightarrow Q_{\rho2}(\omega)-\mat{1}{\coth(\omega/2T_2)}{0}{-1}\nu\Gamma(\omega).
\end{equation}
The corresponding shift in the  $+$-$-$ basis, relevant for the scattering terms in Eq.(\ref{tunnelS}), is
$q^\alpha_{\rho2}(t)\rightarrow q^\alpha_{\rho2}(t)-e^*V_{\rm sd}t-(e^*/\sqrt{2})(P^\alpha\mu)(t)$,
where the effective charge $e^*=\nu e$ and $(P^\alpha\mu)(t)=\int (d\omega/2\pi)[\coth(\omega/2T_2)-\alpha]\mu(\omega)$.

We now compute the effects of the backscattering using standard Keldysh perturbation theory \cite{rammerbook}. After 
implementing the above shift the Keldysh partition function to $\mathcal{O}(\zeta_i^2)$ can be computed as 
$Z^K=Z^K_0\left[1+\langle (S^K_{T})^2\rangle_0\right]$, where 
$Z^K_0=\prod_{mj}\int\mathscr{D}Q_{mj}\mathscr{D}Q^\dagger_{mj}\exp\{S^K_{\rm eff}\}$, and $\langle\cdots\rangle_0$ denotes 
averaging with respect to the weight $\exp\{S^K_{\rm eff}\}$. The steady-state current and the DC component of its noise can then be 
computed by taking standard functional derivatives with respect to the source field $\mu(\omega)$ \cite{kfnoise}
\begin{eqnarray}
I&=&\frac{i}{\sqrt{2}}\int\frac{d\omega}{2\pi}e^{-i\omega t}\left.\frac{\delta\ln Z^K}
{\delta\mu(-\omega)}\right|_{\mu=0},\\
S^{DC}&=&-\lim_{\omega\rightarrow 0}\frac{1}{2}\int\frac{d\omega'}{2\pi}
\left.\frac{\delta^2\ln Z^K}{\delta\mu(\omega')\delta\mu(\omega)}\right|_{\mu=0}.
\end{eqnarray}

In the absence of backscattering the current is simply given by $I_0=\nu e^2V_{\rm sd}/2\pi$. The backscattered current reads
\begin{multline}
\label{backIres}
I_B=\frac{ie^*}{2}\int dt
\left[\frac{\zeta_1^2+\zeta_2^2}{2}\sin\left(\frac{e^*V_{\rm sd}t}{2}\right)+
\zeta_3^2\sin(e^*V_{\rm sd}t)\right]\\\times F(T_1,t)F(T_2,t)
\end{multline}
where $F(x,t)=\left(\pi x\tau_c/\sin\pi x(\tau_c+it)\right)^\nu$ and $\tau_c^{-1}$ is the UV cutoff. Likewise, the noise in the absence of 
backscattering is the usual Johnson-Nyquist term, $S^{DC}_0=2\nu e^2T_2/4\pi$. The correction coming from backscattering is given by
\begin{multline}
S^{DC}_B=\frac{(e^*)^2}{2}\int dt\left[\frac{\zeta_1^2+\zeta_2^2}{4}
\cos\left(\frac{e^*V_{\rm sd}t}{2}\right)+\zeta_3^2\cos\left(e^*V_{\rm sd}t\right)\right]\\
\times F(T_1,t)F(T_2,t)(1-2itT_2).
\end{multline}

The excess noise is defined as $S^{DC}_{\rm ex}(T_1,T_2,V_{\rm sd})=S^{DC}(T_1,T_2,V_{\rm sd})-S^{DC}(T_2,T_2,V_{\rm sd}=0)$,
where $S^{DC}=S^{DC}_0+S^{DC}_B$.
For $V_{\rm sd}=0$, the plot of $S^{DC}_{\rm ex}$ as a function of $\Delta T=T_1-T_2$ is shown in Fig. \ref{fig:noise}. The excess noise 
is plotted as a function of the source-drain voltage, $eV_{\rm sd}$, in Fig. \ref{fig:noisebias}.

The above results can be extended to arbitrary QH states by noting that even for non-abelian QH states \cite{BeNa06} the 
only characteristics of a state which enter the calculation of the current and noise to lowest order in the backscattering strength are 
the QP charge $e^*$ and the local scaling dimension $g$ of the most relevant edge creation operator for QPs $\hat{T}(x,t)$, defined 
via the time decay of the expectation value $\langle \hat{T}^\dagger(x_0,t) \hat{T}(x_0,0)\rangle \sim t^{-g}$. The scaling dimension 
$g$ replaces the exponent $\nu$ in the correlation function $F(x,t)$, and using the appropriate QP charge we find for the excess noise
$S^{DC}_B\propto\zeta^2 \int dt \cos\left(e^* V_{\rm sd}t\right)F(T_1,t)F(T_2,t)(1-2itT_2)$, where $\zeta$ is the tunneling amplitude 
for the backscattering process. There is some theoretical \cite{BiNa09,ReSi09}  and experimental  \cite{Radu+08} evidence that the 
anti-Pfaffian state may be the correct description for the experimentally realized state at filling fraction $\nu=5/2$, and at the random 
fixed point one finds $e^* = 1/4$ and $g=1/2$. The excess noise for the anti-Pfaffian is shown in Figs. \ref{fig:noise} and 
\ref{fig:noisebias}. 

The theoretical results shown in Figs. \ref{fig:noise} and \ref{fig:noisebias} agree well with the experimental ones 
\cite{heiblum} if one makes the identification $T_1- T_2 \propto I_1$. In Fig. \ref{fig:noisebias}, one sees that  the slope of the excess noise 
as a function of $V_{sd}$ decreases with increasing $T_1$. This is in agreement with the experimental result that the quasi-particle charge 
obtained from the slope of excess noise as a function of impinging current decreases with increasing current $I_1$. The experimental finding 
that the current $I_1$ influences the noise for filling fraction $\nu=5/2$ is inconsistent with the Moore-Read state \cite{MR} which has no 
counter-propagating neutral mode. The abelian strong-pairing $K=8$ candidate state \cite{K=8} for $\nu=5/2$ is ruled out because it has 
no neutral mode, and the abelian (331)-state \cite{Halperin83} and the non-abelian $SU(2)_2 \times U(1)$ state \cite{K=8,OvWe08} are ruled
out because they have co-propagating neutral modes. For these reasons, the experiment  \cite{heiblum} indicates that  the  $\nu=5/2$ state
may be described by either the anti-Pfaffian \cite{apf1,apf2} or an edge reconstructed Pfaffian state \cite{Wan+06,OvWe08}. 

We thank A.~Bid, M.~Heiblum, N.~Ofek and A.~Stern for useful discussions. This work was supported by the German Ministry of Education 
and Research Grant No. 01BM0900.

\vspace{-.5cm}


\begin{thebibliography}{999}
\bibitem{halperin} B.I. Halperin, Phys. Rev. B \textbf{25}, 2185 (1982).
\bibitem{buettiker} M. B\"{u}ttiker, Phys. Rev. B \textbf{38}, 9375 (1988).
\bibitem{wen} X.-G. Wen, Phys. Rev. B \textbf{43}, 11025 (1991); Phys. Rev. Lett.
\textbf{64}, 2206 (1990).
\bibitem{macdonald} A.H. MacDonald, Phys. Rev. Lett. \textbf{64}, 222 (1990);
M.D. Johnson and A.H. MacDonald, Phys. Rev. Lett. \textbf{67}, 2060 (1991).
\bibitem{ashoorietal} R.C. Ashoori \textit{et al}., Phys. Rev. B \textbf{45},
3894 (1992).
\bibitem{kfp} C.L. Kane, M.P.A. Fisher and J. Polchinski, Phys. Rev. Lett. \textbf{72},
4129 (1994).
\bibitem{review}  C.~Nayak, S.~H.~Simon,
 and A.~Stern, , M.~Freedman, and S.~Das Sarma, Rev. Mod. Phys. 80, 1083 (2008).
 \bibitem{MR} G. Moore and N. Read, Nucl. Phys. B {\bf 360}, 362 (1991).
\bibitem{apf1} M. Levin, B. I. Halperin, and B. Rosenow, Phys. Rev. Lett. 99, 236806 (2007).
\bibitem{apf2} S.-S. Lee, S. Ryu, C. Nayak, and M. P. A. Fisher, Phys. Rev. Lett. 99, 236807 (2007).
\bibitem{OvWe08} B. J. Overbosch and X.-G. Wen, arXiv:0804.2087 (2008).
\bibitem{heiblum} A. Bid, N. Ofek, H. Inoue, M. Heiblum, C.L. Kane, V. Umansky, and D. Mahalu, arXiv:1005.5724 (2010). 
\bibitem{Granger+09} G. Granger, J.P. Eisenstein, and J.L. Reno, Phys. Rev. Lett. {\bf 102}, 086803 (2009). 
\bibitem{FeLi08} D.E. Feldman and F. Li, Phys. Rev. B {\bf 78}, 161304 (2008).
\bibitem{GrDa09} E. Grosfeld and S. Das, Phys. Rev. Lett. {\bf 102}, 106403 (2009). 
\bibitem{Wan+06} X. Wan, K. Yang, and E. H. Rezayi, Phys. Rev. Lett. 97, 256804 (2006).
\bibitem{ChHa} D.B. Chklovskii and B.I. Halperin, Phys. Rev. B {\bf 57}, 3781 (1998).
\bibitem{kfnm} C.L. Kane and M.P.A. Fisher, Phys. Rev. B \textbf{51}, 13449 (1995).
\bibitem{kfnoise} C.L. Kane and M.P.A. Fisher, Phys. Rev. Lett. \textbf{72}, 724 (1994).
\bibitem{kamenev} A. Kamenev, in \textit{Nanophysics: Coherence and Transport}, Ed. H. Bouchiat 
(Elsevier, New York, 2005), pp. 177 - 246.
\bibitem{rammerbook} J. Rammer, \textit{Quantum Field Theory of Nonequilibrium States}, (Cambridge, 
Cambridge, 2007).
\bibitem{BeNa06} C. Bena and C. Nayak, Phys. Rev. B 73, 155335 (2006). 
\bibitem{BiNa09} W. Bishara and C. Nayak
Phys. Rev. B 80, 121302 (2009).
\bibitem{ReSi09} E.H. Rezayi and  S.H. Simon, preprint arXiv:0912.0109 (2009). 
\bibitem{Radu+08}   L.P. Radu J.B. Miller, C.M. Marcus, M.A. Kastner,
L.N. Pfeiffer, and K.W. West, Science 320, 899 (2008).
\bibitem{Halperin83} B. I. Halperin, helv. phys. acta {\bf 56}, 75 (1983).
\bibitem{K=8} X. G. Wen, Phys. Rev. Lett. 66, 802 (1991); B. Blok and X. G. Wen, Nucl. Phys. B 374, 615 (1992).




\end{thebibliography}
\end{document}